# Comment on "Localization and the Mobility Edge in One-Dimensional Potentials with Correlated Disorder"

In a recent Letter [1], Izrailev and Krokhin derived an equation for the wave-function localization length $l$ in terms of two-point correlation function of a one-dimensional (1D) weak random potential, and used it to argue for the existence of mobility edges for certain types of correlated disorder in 1D. The purpose of this Comment is twofold: first, to present a more general derivation of the same equation using the the standard weak-localization theory, and second, to correct the mistake in the algorithm for the construction of a given correlated random potential proposed in Ref.1. A direct numerical calculation of $l$ in a correlated random potential generated by my improved procedure is then found to be in perfect agreement with the analytic expression, thus removing the discrepancy between the two detected in Ref. 1.

In a weak random potential $\epsilon(x)$, the leading correction to Boltzmann conductivity of a 1D electronic system $\sigma_B = e^2 v_F \tau/(\pi\hbar)$ (where $v_F = dE(k)/dk|_{k_F}$ is the Fermi velocity, $\tau$ backward scattering time, and $e$ charge) is $\sigma_Q = -e^2 L/(\pi\hbar)$, where $L$ is the length of the system [2]. Balancing the two terms gives $l = v_F \tau$ in 1D, as well-known. On the other hand, if $\langle \epsilon(x)\epsilon(y)\rangle = \epsilon_0^2 \xi(x-y)$, in Born approximation $\tau^{-1} = \epsilon_0^2 \tilde{\xi}(2k_F)\pi\mathcal{N}_F$, in 1D, where $\mathcal{N}_F$ is the density of states at the Fermi level, and $\tilde{\xi}(k)$ the Fourier transform of $\xi(x)$. For the tight-binding dispersion $E(k) = -2\cos k$ one thus finds $\tau^{-1} = \epsilon_0^2 \tilde{\xi}(2k_F)/(4\sin(k_F))$, which together with $v_F = 2\sin k_F$ gives finally

$$l^{-1} = \frac{\epsilon_0^2 \tilde{\xi}(2k_F)}{8\sin^2 k_F}. \qquad (1)$$

This is precisely the Eq. 12 in ref. 1, upon identification of the parameter $\mu$ with the Fermi wave vector $k_F$.

The above equation implies that a weak random potential with $\tilde{\xi}(2k_F) = 0$ will produce extended states at the Fermi energy, simply because backscattering process is absent. To independently check this proposition one may construct a correlated random sequence $\{\epsilon_i\}$ with a given two-point correlator $\langle \epsilon_n \epsilon_m\rangle = \epsilon_0^2 \xi(|m-n|)$ as

$$\epsilon_n = \sum_{m=-\infty}^{\infty} r_m c(m-n), \qquad (2)$$

where $\langle r_m r_n\rangle = \epsilon_0^2 \delta_{m,n}$. Evidently, then $\langle v_n v_m\rangle = \epsilon_0^2 \sum_j c(n-j)c(j-m)$, or

$$\tilde{c}(k) = \tilde{\xi}^{1/2}(k), \qquad (3)$$

where $\tilde{f}(k) = \sum_n f(n)\cos(kn)$, is the discrete Fourier transform. Instead of Eq. 3, Izrailev and Krokhin assumed that $c(n) = \xi(n)$ (their Eq. 19). This does not matter if $\tilde{\xi}(k)$ is a step function, and the authors correctly reproduced the discontinuous drop of $1/l$ at the mobility edge. For $\tilde{\xi}(2k) = 2(\cos k - \cos(\pi/3))$ for $0 < k < \pi/3$, $\tilde{\xi}(2k) = 0$ for $\pi/3 < k < 2\pi/3$, and $\tilde{\xi}(2k) = 2(\cos(2\pi/3) - \cos k)$ they noticed, however, that $1/l$ vanishes faster than linearly near $E = 1$, in contradiction to the Eq. 1. This was interpreted as being an artifact of retaining only a finite number of terms in the sum in Eq. 2. On Fig. 1 I exhibit the result of the numerical calculation on a discrete system of $M = 10^6$ sites, using the correlated potential constructed from Eqs. 2 and 3. Here $l^{-1} = M^{-1}\sum_{n=1}^{M-1} \ln(|\Psi_{n+1}/\Psi_n|)$, and $-\Psi_{n+1} + \epsilon_n \Psi_n - \Psi_{n-1} = E\Psi_n$. Infinite sum in Eq. 2 is approximated with first 100 terms, and $r_i$ are chosen randomly from the interval $[-0.1, 0.1]$. Mobility edge at $E = 1$ and the linear behavior of $l^{-1}$ in its vicinity are well reproduced. I also checked that omitting the power $1/2$ in Eq. 3 indeed leads to faster than linear dependence of $l^{-1}$ similar to the one found in Ref. 1.

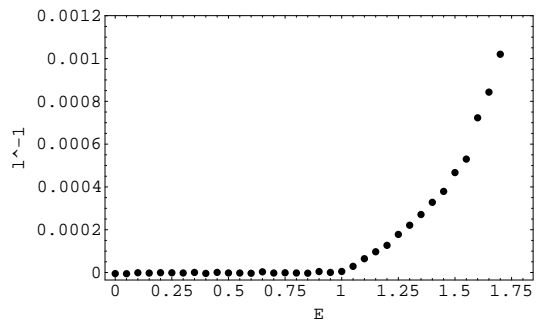

FIG. 1. Inverse of the localization length $l$ as a function of energy E in a specific correlated random potential.

Finally, the algorithm for generation of correlated random potentials presented here can be straightforwardly generalized to higher dimensions.

Igor F. Herbut
Department of Physics
Simon Fraser University
Burnaby, BC V5A 1S6, Canada

1